\documentclass[twocolumn, superscriptaddress, prapplied]{revtex4-2}

\usepackage{graphicx}
\graphicspath{{./Figures/}}
\usepackage{hyperref}
\hypersetup{backref,colorlinks=true,allcolors=blue}

\usepackage{float,amsmath,amssymb,multirow,tabularx, tikz}

\usetikzlibrary{shapes,arrows,calc}
\usepackage[caption=false]{subfig}
\usepackage[most]{tcolorbox}

\usepackage{tikz}
\usetikzlibrary{calc,arrows,shapes,positioning}
\usetikzlibrary{patterns,snakes}
\usetikzlibrary{positioning,decorations.pathreplacing}

\definecolor{sangria}{rgb}{0.57, 0.0, 0.04}
\definecolor{arsenic}{rgb}{0.23, 0.27, 0.29}
\definecolor{prussianblue}{rgb}{0.0, 0.19, 0.33}
\definecolor{phthalogreen}{rgb}{0.07, 0.21, 0.14}
\definecolor{dgreen}{rgb}{0.0, 0.4, 0.2}

\tikzstyle{block1} = [rectangle, draw=arsenic!50, fill=arsenic!20, 
    text width=10em, text centered,  minimum height=3em,line width=1mm]
\tikzstyle{block2} = [rectangle, draw=sangria!50, fill=sangria!20, 
    text width=10em, text centered,  minimum height=3em,line width=1mm]    
\tikzstyle{block3} = [rectangle, draw=prussianblue!50, fill=prussianblue!20, 
    text width=10em, text centered, minimum height=3em,line width=1mm]    
\tikzstyle{block4} = [rectangle, draw=phthalogreen!50, fill=phthalogreen!20, 
    text width=10em, text centered, minimum height=3em,line width=1mm]    
    
\tikzstyle{line} = [draw=arsenic, -latex', line width=0.1cm]

\def\td#1{$_\mathrm{#1}$}
\def\tu#1{$^\mathrm{#1}$}

\def\ep{\mathcal{E}}
\def\td#1{$_\mathrm{#1}$}
\def\tu#1{$^\mathrm{#1}$}
\def\der#1{d\mathbf{#1}\;}
\def\im{\mathit{i}}
\def\gw{G$_0$W$_0$ }

\def\Im#1{\mathfrak{Im}\left[#1\right]\;}

\def\fig#1{Fig.~\ref{#1}}

\newcommand{\splitbox}[1]{%
  \linespread{1}\selectfont
  \renewcommand*{\arraystretch}{1.5}%
  \begin{tabular}{@{}c@{}c@{}}
    \strut#1\strut\strut
  \end{tabular}%
  }

\DeclareUnicodeCharacter{2212}{-}
\DeclareUnicodeCharacter{2009}{\,} 

\begin{document}

\author{Okan K. Orhan}
\affiliation{School of Physics, SFI AMBER Centre and CRANN Institute, Trinity College Dublin, The University of Dublin, Dublin 2, Ireland}
\affiliation{Department of Mechanical Engineering, University of British Columbia, Vancouver, BC, V6T 1Z4, Canada}

\author{Frank Daniel Bello}
\affiliation{School of Physics, SFI AMBER Centre and CRANN Institute, Trinity College Dublin, The University of Dublin, Dublin 2, Ireland}
\
\author{Nicol\'{a}s Abad\'{i}a}
\affiliation{School of Physics and Astronomy and Institute for Compound Semiconductors, Cardiff University, Cardiff, CF24 3AA, U.K}

\author{Ortwin Hess}
\affiliation{School of Physics, SFI AMBER Centre and CRANN Institute,  Trinity College Dublin, The University of Dublin, Dublin 2, Ireland}

\author{John F.  Donegan}
\affiliation{School of Physics, SFI AMBER Centre and CRANN Institute,  Trinity College Dublin, The University of Dublin, Dublin 2, Ireland}
\author{David D. O'Regan}
\affiliation{School of Physics, SFI AMBER Centre and CRANN Institute,  Trinity College Dublin, The University of Dublin, Dublin 2, Ireland}

\title{Multiscale Design of Au-Based Alloys for Improved Plasmon Delivery and Nanoheating in Near-Field Transducers}

\begin{abstract} 
Plasmonic near-field transducers (NFTs) play a key role in administering nanoscale heating for a number of applications ranging from medical devices to  next generation  data processing technology. We present a novel multi-scale approach, 
combining quantum many-body perturbation theory with
finite-element modelling, 
 to predict the electric and thermal material parameters of various Au-based, noble metal (M) alloys.
Specifically, we focus on modelling their performance within an NFT designed to focus high-intensity, sub-diffracted light for technologies such as nanoscale etching, manipulation, sensing, and heat-assisted magnetic recording (HAMR).
Elemental Au is the long-standing general-purpose NFT medium due its excellent plasmonic performance at  relevant wavelengths. 
However, elemental Au is a soft, ductile material that tends to  extrude and deform in response to  extreme temperature gradients.
Therefore, alloying Au with other noble metals  such as Ag, Cu, Pd or Pt, has attracted considerable interest for improved mechanical and thermal robustness while reaching threshold plasmonic generation at standard optoelectronics operating wavelengths (e.g., $\approx 830$~nm) and approximate high-power NFT temperatures ($\approx 400$~K). 
We predict that certain Au-Ag alloys %, particularly ordered-phase AuAg\td{3} and solid solution Au\td{0.125}Ag\td{0.875}, 
may offer improved thermal stability as whole-NFT media compared to elemental Au,   
alongside plasmonic figures of merit comparable to that of Au.
Simulations of certain solid solution Au-Pd/Pt alloys enable us to predict significantly enhanced thermal conductivity. %,
%but that this does not imply improved performance 
%when these are used as whole-NFT media.
%
We  predict %, however, 
that alloying with Pd at low concentrations $\sim 10\%$ 
may preserve the NFT performance of Au, 
while offering the benefits of improved thermal and mechanical stability.

\end{abstract}

\maketitle
\section{Introduction}\label{sec:s1}

Au nano-particles  and thin-films are widely used in plasmonic media for opto-electronic applications such as optical and bio-sensing~\cite{Alagiri2017,bengali2018gold,RODRIGUES201874}, light and heat-assisted drug delivery systems~\cite{JAIN200718,Zhang2015,Tran2017},  as well as data processing and recording~\cite{Bello2020,Bello2019,black2000magnetic,Mansuripur:09,sato2009heat,zou2014recording}.
This is  due to their high plasmonic performance  throughout the near-infrared-visible (IR-vis) spectral range~\cite{Amendola_2017}.
However, these applications often operate at temperatures above room temperature, where dephasing rates increase and the thermal conductivity of Au decreases significantly. This can be detrimental for nanoscale components and device operation~\cite{doi:10.1063/1.4817582}.
It is well known that Au is soft and may protrude or deform regardless of design and given increases in temperature as little as a few tens of Kelvin~\cite{Abbott2019}.
These issues greatly reduce the thermal stability and therefore manufacturability of many Au-based nano-devices.
Hence, alloying Au with other noble metals such as Ag, Cu, Pd and Pt has been proposed to engineer both plasmonic and thermal properties for designated applications~\cite{Blaber_2010,ADOM:ADOM201500446,PhysRevB.6.1209,Nishijima2014,doi:10.1021/acsphotonics.5b00586,hashimoto2016ag}.
This is  true, for example,  in heat-assisted magnetic recording (HAMR), 
where the temperatures of near-field transducers (NFTs) may exceed 400 K~\cite{Datta2017,Bello:20}. 

This present work is a proof-of-concept for a  
multi-scale modelling workflow that goes from beyond-density-functional theory 
first-principle electronic structure upwards in scale.
Using this, we carried out a simulated performance assessment of Au-based alloys 
as they fully or partially substitute the metallic components within a metal-insulator-semiconductor (MIS) NFT 
that was previously designed to be used in the HAMR device, as described in detail in Ref.~\citenum{Abadia:18}.
The approach developed, and the alloys explored, could
in principle also be used to improve other NFT designs used in 
nano-heating applications such as bowtie, E- or C-aperture, as well as a number of nanoparticle designs such as nanoparticle-on-mirror (NPoM) structures, to name a few.
Since it is impractical to explore all possible alloy compositions, in what follows
in this proof of principle will look at two different alloying regimes, 
 focusing first on evaluating the  plasmonic performance of mono-layer-stacked and fully disordered bulk Au-Ag/Cu alloys, followed by that of fully 
disordered bulk Au-Pd/Pt alloys, throughout their composition space.
The multi-scale approach involves calculating first-principles electronic
parameters including quasiparticle renormalizaiton effects at the 
perturbative G$_0$W$_0$ level, and on this basis then generating 
semi-empirical plasmonic and thermal material parameters,
which in the final step are fed into finite-element method 
(FEM) simulations to assess NFT device-level performance of the simulated alloys. 

The example NFT design that we use as an example here~\cite{Abadia:18} is from the context of HAMR, a 
next-generation  high-density recording technology expected to 
store information on a ground-breaking scale of tens of terabytes per square 
inch~\cite{black2000magnetic, zou2014recording}. 
The fundamental principle behind HAMR is the nanosecond-scale heating of a local region in the recording medium above its Curie temperature using sub-diffraction focused light; and by doing so, the reducing of the coercivity of the region in
order to enable magnetic recording. 
The key component of this operation is  the near-field transducer~\cite{Datta2017,YANG2017159} operating as a plasmonic antenna used for local and fast heating.  
Depending on the NFT, such a task requires either a localized surface-plasmon (LSP)~\cite{1367-2630-4-1-393,doi:10.1021/jp0361327}, or a symmetric surface plasmon polariton (SPP)~\cite{ZAYATS2005131} mode at the operation wavelength.
Regardless of the particular design, the heat-write-cool operation cycle has to  be safely repeated many thousands of times for a viable HAMR device.  
For this, a sufficiently high plasmon population is required to heat the recording medium above its Curie temperature, while avoiding overheating of the metallic tip and accompanying heat sink in NFTs at the operation wavelength, for which we choose $830$~nm following Ref.~\citenum{Abadia:18}.
\begin{figure}[H]
\centering
\includegraphics[width=0.5\textwidth]{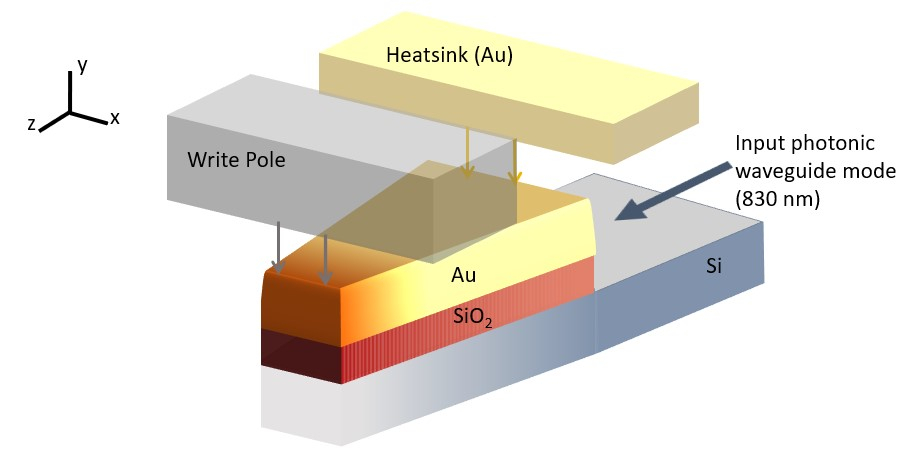}
\caption{Schematic diagram of the hybrid antenna-based NFT 
used in multi-scale simulations, as designed in Ref.~\citenum{Abadia:18},
demonstrating the placement of the write pole and metallic heat-sink that is composed of Au or an Au alloy. 
The air-bearing surface (ABS) lies in front of the narrowest tapered region, 
which sits only a few nanometers above the recording medium.
Coupled FEM simulations of Maxwell's equations along with the heat equation were carried out,~\cite{Bello:20,Abadia:18,comsol} with full simulation details provided in the Supplemental Material~\cite{SeeSM}.}
\label{fig:nft-design}
\end{figure}
The  NFT design in question, schematically illustrated in Fig.~\ref{fig:nft-design}, is a plasmonic antenna designed as a MIS layered structure that supports a SPP in order to produce sub-diffraction focused light, which is then converted to heat within the recording medium.
This NFT was designed by Abad\'{i}a \textit{et al.} for SPP generation at a single mode wavelength of $830$~nm, particularly for operation in a HAMR apparatus~\cite{Abadia:18,7296641,Zhong:18}. 
For optimal HAMR operation, heating should be confined at a  $50\times  50$~nm$^2$ area or less on the recording medium. 
This is achieved by using a plasmonic tip as small as  $20$~nm in the cross-track (x) direction of Fig.~\ref{fig:nft-design}~\cite{Abadia:18}. 
The small size scale combined with the softening tendency of the elemental Au under relatively high temperatures makes the plasmonic tip potentially 
vulnerable to deformation. 
Therefore, it would be beneficial if the Au component of the nanostructure can be replaced with alloys that could significantly increase the thermal stability and hardness of the NFT, in particular at the air-bearing surface~\cite{Abadia:18,7296641}. 
\begin{figure}[t]
\centering
\tcbox[sharp corners, boxsep=0.0mm, boxrule=0.5mm,  colframe=gray, colback=white]{
\resizebox{0.4\textwidth}{!}{
\begin{tikzpicture}
    % Place nodes
    \node [block1] (vca) {VCA};
    \node [block1, below of=vca, node distance=2.5cm] (ksdft) {KS-DFT};
    \draw [line,line width=0.5mm] (vca) -- (ksdft) node [midway, left] (TextNode) {PsP};  
 
    \node [block1, left of=ksdft,node distance=7.5cm] (g0w0) {\gw};

    \draw [line,line width=0.5mm] (ksdft.north west) -- (g0w0.north east) node [midway, above] (ksout) {};%\Large $E_i^{DFT}, \psi_i^{DFT}$};
     \draw [line,line width=0.5mm] (g0w0.south east) -- (ksdft.south west) node [midway, below] (g0w0out) {\Large $s_\mathrm{v}^\mathrm{QP},s_\mathrm{c}^\mathrm{QP}$};
   
    \node [block1, below  of=ksdft,node distance=5cm] (elpar) {Electronic parameters};
   \node [block2, below  of=g0w0,node distance=5cm] (oppar) {Optical parameters};
   
   \node [block2,  below  of=g0w0out,node distance=2cm] (plapar) {Plasmonic parameters};

   \node [block2,  below  of=plapar,node distance=5cm] (thepar) {Thermal parameters};
   \node [block2,  below of=thepar,node distance=2.5cm] (finel) {Finite-element simulations};

   % Draw edges

   \draw [line,line width=0.5mm] (g0w0out) -| (oppar) node [above] (TextNode) {};
   \draw [line,line width=0.5mm] (g0w0out) -| (elpar) node [above] (TextNode) {};
   \draw [line,line width=0.5mm] (g0w0out) -- (plapar) node [above] (TextNode) {};
   \draw [line,line width=0.5mm] (plapar) |- (oppar) node [above] (TextNode) {};
   \draw [line,line width=0.5mm] (elpar) |-  (thepar) node [above] (TextNode) {};
   \draw [line,line width=0.5mm] (plapar) -- (thepar) node [above] (TextNode) {};
    
  \draw [line,line width=0.5mm] (elpar) |-  (finel) node [above] (TextNode) {};
  \draw [line,line width=0.5mm] (thepar) -- (finel) node [above] (TextNode) {};
  \draw [line,line width=0.5mm] (oppar) |-  (finel) node [above] (TextNode) {};

% \draw [decorate, decoration={ brace, raise=1cm ,amplitude=10pt}] (thepar.east) -- (g0w0.west) node [pos=0.5,anchor=north, yshift=-1.5cm] (qp) {};
%  
%  %\draw [decorate, decoration={ brace, raise=1cm ,amplitude=10pt}] (elpar.east) -- (plapar.west);
%  %\draw [decorate, decoration={ brace, raise=4cm ,amplitude=10pt}] (elpar.east) -- (oppar.west);

\end{tikzpicture}}}
\caption{Schematic illustration of the work-flow followed here to obtain the material parameters for  finite-element simulations. The virtual crystal approximation (VCA) is used to model virtual atoms of Au alloys with pseudo-potentials (PsP), which are then fed into the approximate Kohn-Sham density-functional theory (KS-DFT). Band stretching factors 
($s^{QP}_\mathrm{v}$
and $s^{QP}_\mathrm{c}$) are then obtained by means of linear regression to 
 quasiparticle energies calculated using one-shot non-self-consistent 
many-body perturbation theory
\ (G$_0$W$_0$). Please see Section \ref{sec:s2} for details. 
Color codes are assigned to the first-principles parameters as gray and the the semi-empirical parameters as pink. See the Supplemental Material~\cite{SeeSM} for full lists of parameters used and details of the computational methodology.}
\label{fig:wf}
\end{figure}
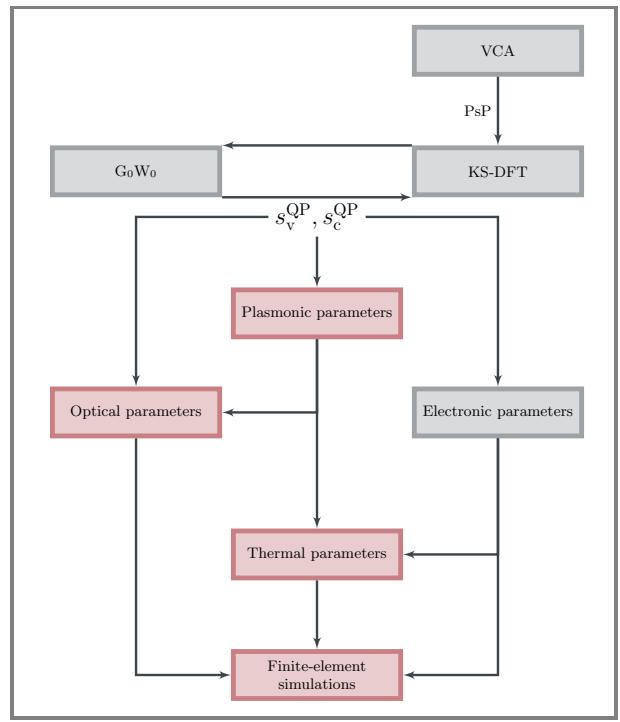

The primary criterion for any NFT design is generating sufficient plasmon populations, particularly high-quality SPPs for this particular design. 
Two plasmonic figure-of-merits (p-FOMs) are used to assess 
the quality of SPP generation in this work.
The first p-FOM is the electron-energy loss spectrum  (EELS) given by~\cite{Boersch1962,raether1965electron,Froitzheim1977,PhysRev.101.554,PhysRev.82.625,PhysRev.85.338, PhysRev.92.609} 
\begin{align}\label{eq:eq1}
\mathrm{EELS}=-\Im{1/\ep},
\end{align} 
where $\ep=\ep_1+\im \ep_2$ is the complex macroscopic dielectric function. 
EELS checks for a sufficient presence of nearly free electrons available for collective oscillations rather than forming particle-hole pairs at a given wavelength~\cite{ROCCA19951}. 
The second p-FOM  is the quality of a generated SPP, indicating its longevity, 
given by~\cite{0953-8984-22-14-143201} 
\begin{align}\label{eq:eq2}
Q_{\mathrm{SPP}}(\omega)=\ep_1^2/\ep_2.
\end{align} 
We obtained the macroscopic dielectric function using the procedure defined in Ref.~\citenum{Orhan_2019} along with thermal parameters such as thermal conductivity and heat capacity. %
Furthermore, we extracted the relevant quantities in the cases of  ordered Au-Ag and Au-Cu alloys directly from the same work.
This method is summarized in Section \ref{sec:s2} and the description of results and optimum Au-alloys for nano-heating is provided in Section \ref{sec:s3}.

\section{Theoretical Spectroscopy Simulations}\label{sec:s2}
A systematic illustration of the work-flow used 
to obtain the material parameters needed for the finite-element 
simulations is shown in \fig{fig:wf}. 
Approximate Kohn-Sham density-functional theory (KS-DFT)~\cite{PhysRev.136.B864,PhysRev.140.A1133}, which uses local and semi-local approximations for exchange and correlation~\cite{PhysRev.140.A1133,PhysRevB.21.5469,PhysRevB.33.8822,PhysRevB.46.6671}, 
was used for constructing an initial approximation for the band-structure.
KS band-structures, in particular, offer convenient starting points for 
perturbative linear-response theoretical spectroscopy approaches, 
such as the random-phase approximation (RPA)~\cite{PhysRev.82.625,PhysRev.85.338,PhysRev.92.609,RevModPhys.36.844}. 
However, the next accuracy of the RPA is highly dependent 
on the accuracy of electronic band-structure provided as its starting point, 
and in particular the dressing of independent particles and holes by
excitations can be important.
It has been shown that KS-DFT using local-density approximations performs poorly in the case of noble metals~\cite{PhysRevB.86.125125,Orhan_2019} due to absent 
non-local electronic exchange-correlation effects.
This error becomes more pronounced in spectral simulations, not least in the case of noble metals~\cite{0305-4608-14-1-013,doi:10.1021/cr200107z,PhysRevLett.88.016403,PhysRevB.66.115101,PhysRevB.66.161104,Orhan_2019}.
This can be partially overcome by providing a quasi-particle band-structure,
rather than the non-interacting Kohn-Sham DFT band-structure, 
as the starting point for RPA. 
In this work, we applied a set of stretching operators, which were obtained  using one-shot, non-self-consistent GW (G$_0$W$_0$)~\cite{MARINI20091392,Orhan_2019} from first principles to the  KS band-structures. 
The resulting stretched KS band-structures are used both in spectral simulations and determination of fundamental material parameters such as thermal conductivity (See Supplemental Material~\cite{SeeSM} for details).

The conventional RPA operates in the product space of the occupied and unoccupied KS wave-functions, and these occupancies need to be defined with high precision.
In metals,  in order to accurately recover the Drude peak
generated by intra-band transitions within the RPA, 
an unfeasibly dense Brillouin zone sampling of the Fermi surface
is required~\cite{marini2001optical}.
As a result, and following established practice in this area, 
in this work the Drude plasmon  contribution 
(also called the intra-band contribution)  to the macroscopic dielectric function ($\mathcal{E}$) is included via the Drude-Lorentz classical model~\cite{ANDP:ANDP19003060312,ANDP:ANDP19003081102,lorentz1909theory,fox2002optical}, where
it is given by
\begin{align}
\mathcal{E}^{\mathrm{intra}}(\omega)=
\varepsilon_{\infty}-\frac{\omega_\mathrm{p}^2}{\omega^2-\im \eta_\mathrm{p} \omega}.
\end{align}
Here, $\omega_\mathrm{p}$ , $\eta_\mathrm{p}$, and $\varepsilon_{\infty}$ are the Drude plasmon energy,  the phenomenological inverse life-time, and the electric permittivity at the infinite-frequency limit,  respectively. 
In order to construct a practical multi-scale work-flow appropriate to 
materials exploration and interpolation, 
a phenomenological inverse life-time and the electric 
permittivity are determined from the experimental 
spectra of the elemental metals in the infra-red region.  
For a uniform, non-interacting electron gas, the Drude plasmon energy can be approximated  by~\cite{ANDP:ANDP19003060312,ANDP:ANDP19003081102,PhysRevB.16.5277}
\begin{align}
\omega_\mathrm{p}^2=\frac{4\pi N(E_\mathrm{F})}{m_{\mathrm{eff}}},  
\end{align}
where  $N(E_\mathrm{F})$ is  the density of states (DOS) at  the Fermi level and $m_{\mathrm{eff}}$ is the 
electron effective mass.
The Drude plasmon energy can, however, be calculated within first principles starting from the electronic band structure using one-band theory~\cite{doi:10.1080/14786435808237011}.
Within the one-band theory, the effective mass of the metallic band is defined by assuming that  these bands have a parabolic dispersion to the Fermi surface given by~\cite{doi:10.1080/14786435808237011},
\begin{align}
m_{\mathrm{eff}}^{-1} &{}\approx  \frac{1}{3} \langle v^2(E_\mathrm{F})\rangle 
\nonumber \\ &{}=
\frac{1}{3}
 \Big(\sum_i\int_{S_\mathrm{F_i}}\der{k} v^2_i(\mathbf{k}) 
 \Big)
 \Big( \sum_j \int_{S_\mathrm{F_j}}\der{k'} \Big)^{-1},  \nonumber \\
\mbox{if} \quad
v^2_i(\mathbf{k})&{}=
\left|\frac{\partial E_{i,\mathbf{k}}}{\partial \mathbf{k}}\right|^2,
\end{align}
where  $S_\mathrm{F_i}$ is  the Fermi surface
of the $i^\textrm{th}$ metallic band. 
This yields a Drude plasmon energy  that can  be approximated as
\begin{align}
\omega_\mathrm{p}^2=\frac{4\pi}{3} N(E_\mathrm{F})\langle v^2(E_\mathrm{F}) \rangle.
\end{align}

Our work finds that Au-Ag binary alloys form continuous face-centered cubic (FCC)  solids for any given stoichiometric ratios, rather than an ordered alloy. 
This is because they exhibit only short-range ordering~\cite{Okamoto1983ag}. 
While the Au-Cu alloys also prefer the continuous FCC phase, the  Au\td{3}Cu, AuCu, and AuCu\td{3} shows ordered structures at $\approx 400-700$~K~\cite{Okamoto1987cu}.
Similarly, Au\td{3}Pd and AuPd\td{3} at $\approx 900-1110$~K, as well as AuPd at $\approx 300-400$~K, show ordered structures with the continuous FCC phase being a general trend for other stoichimetric ratios and temperature ranges~\cite{Okamoto1985pd}.
In the case of the Au-Pt alloys, there is a large miscibility gap, where Au and Pt forms local sub-structures for $\gtrapprox \%15$ ratio of Pt at the lower temperatures; however, there is still the continuous FCC phase of the low-concentration Pt alloys at $\approx 400$~K, and  high-concentration Pt alloys exist in a very narrow temperature belt between the miscibility and liquid phase regions of higher temperatures~\cite{Okamoto1985pt}.
Hence, most of the Au-M alloys in this work have continuous FCC solid solutions without long-range ordering. 
The most  suitable approach would be getting statistical averaging of every possible spatial configuration within a quite large cell, namely the super-cell approximation, to be able to include the disorder in such alloys.
However, it is not feasible for high-throughput simulations. 
An expedient approach, instead, is to replace the Au-M atom pairs with a virtual atom, which interpolates the behaviours of original atoms, namely the virtual crystal approximation (VCA)~\cite{doi:10.1002/andp.19314010507,PhysRevB.61.7877}.
This can be applied within the approximate KS-DFT via  a mixing scheme for the pseudo-potential (PsP) of the virtual atom using the constituting atoms A and B, given by~\cite{PhysRevB.61.7877}
\begin{align}
V_\mathrm{PsP}^\mathrm{virtual}(\mathbf{r},\mathbf{r}')=(1-x)V^\mathrm{A}_\mathrm{PsP}(\mathbf{r},\mathbf{r}')+xV^\mathrm{B}_\mathrm{PsP}(\mathbf{r},\mathbf{r}').
\end{align}
Here, `$\it{x}$' refers to the amount of alloyed metal. The VCA provides a simple scheme for fully disordered case, which  excludes any  ordering, both long-range and short-range.

\begin{figure}[H]
\centering
\includegraphics[width=\columnwidth]{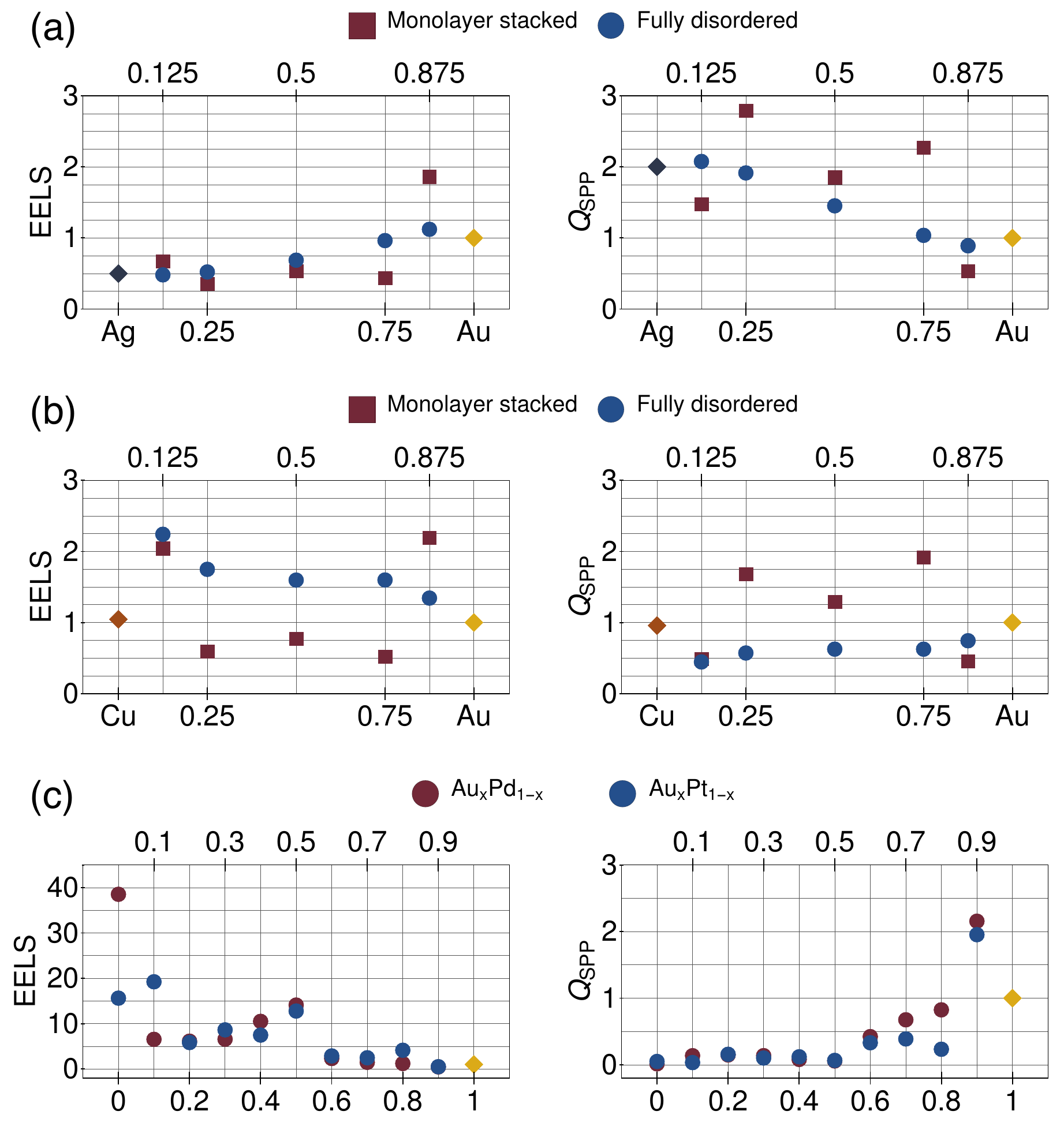}%\vspace{-0.2cm}

\caption{Amplitudes  at $830$~nm of  electron energy-loss (left column) and the surface-plasmon polariton quality factor (right column) of row (a) the mono-layer-stacked and fully disordered Au-Ag alloys; row (b) mono-layer-stacked and fully disordered Au-Cu alloys, and (c)  fully disordered Au-Pd/Pt alloys.
All values are normalized against the elemental Au values, which are set to a value of 1.
The ordered Au-Ag and Au-Cu alloy cases were extracted from Ref.~\citenum{Orhan_2019}, for which the matching fractional stoichiometric ratios for the fully ordered were chosen. 
}
\label{fig:Fig3}
\end{figure}

\section{Results}\label{sec:s3}
\begin{figure*}[t]
\centering
\includegraphics[width=6in]{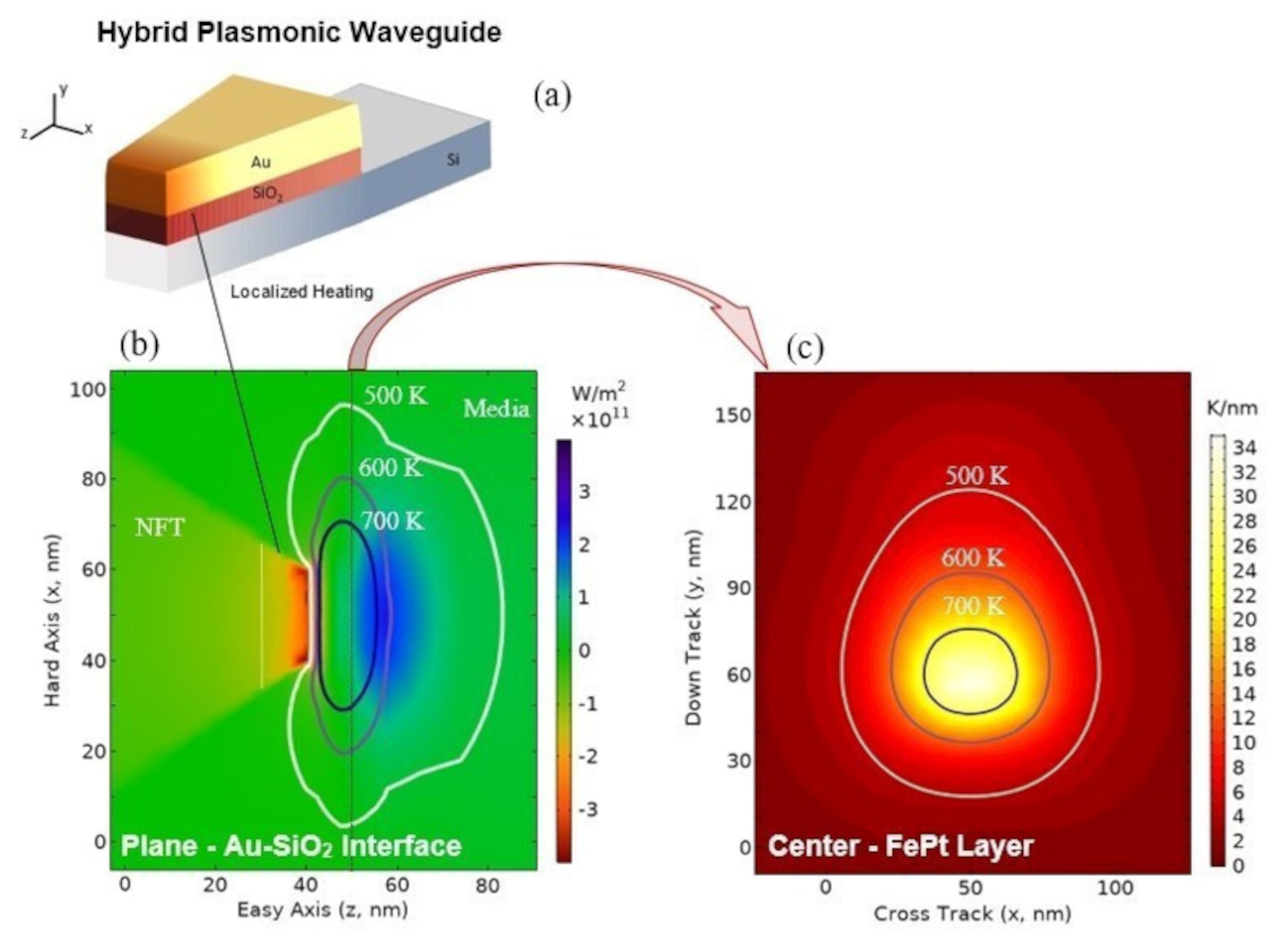}
\caption{ (a) Minimal model of a M-I-S hybrid plasmonic waveguide suitable as an NFT for HAMR, from Ref.~\cite{Abadia:18}. The air-bearing surface (ABS)  is  20 nm along the x-direction (cross track) and 70 nm (Au + SiO\td{2}) in the y-direction (down track). See Supplemental Material~\cite{SeeSM} for a full list of dimensions. (b) Planar cross-section taken from the metal-insulator interface showing the location of maximum heat flux (W/m\tu{2}). AuPt/Pd alloys with concentrations of Pd/Pt of 30, 50, or 70\% were simulated in the NFT region with highest power per unit area (demarked by white dotted line) where softening of the Au may occur given temperatures circa 400 K. (c) Cross section from the center of the recording layer (black vertical line in (b)) located 13.5 nm away from the ABS, showing temperatures reaching above typical values for the Curie temperature of FePt. The magnitude of the temperature gradient is also plotted with maximum cross- and down-track values listed in Table~\ref{tab:therm-perf}.}   
\label{fig:hybrid-plas-wg}
\end{figure*}
Despite being a well-known plasmonic metal,  elemental Ag produces weaker plasmons than Au at $830$~nm,  indicated by its normalized EELS amplitude in Fig.~\ref{fig:Fig3} (a, left), as it has a single and narrow plasmonic peak at $\sim 330$~nm~\cite{Orhan_2019}. 
However, the small population of SPPs of Ag-rich alloys is longer lived compared to that
of elemental Au, shown in Fig.~\ref{fig:Fig3} (a, right). 
Interestingly, the mono-layer-stacked AuAg\td{3} exhibits exceptionally high quality SPPs, while still exhibiting significant simulated generation. 
Furthermore, it has a predicted thermal conductivity of $\kappa=669$~(Wm\tu{-1}K\tu{-1}), significantly higher than 
the $\kappa=312$~(Wm\tu{-1}K\tu{-1}) of the elemental Au. To calculate the conductivity, we included various approximated scattering processes such as the electron-phonon, the electron-electron, as well as the electron-impurity scattering and also subsequently calculated the specific heat capacity for FEM simulations of the heat equation (see Supplemental Material~\cite{SeeSM} for details and values).

The elemental metals Au and Cu exhibit rather similar affinities for SPP generation. 
Disordered Au\td{x}Cu\td{1-x} alloys produce lower quality SPPs, as shown in Fig.~\ref{fig:Fig3} (b), regardless of the fractional ratios.
On the other hand,   mono-layer-stacked Au\td{3}Cu, AuCu, and AuCu\td{3} show improved qualities in simulation, with  strong plasmons as well as improved thermal conductivities compared to both the elemental Au and Cu (see Supplemental Material~\cite{SeeSM}).
As shown in Fig.~\ref{fig:Fig3} (c), Pd- and Pt-rich alloys produce  strong, but quite short-living plasmons. 
On the contrary, Au-rich fully disordered Au-Pd/Pt improve the SPP qualities significantly. 
Unfortunately, mixing Au with either Pd or Pt appears to substantially reduce thermal conductivity, 
 unsurprisingly as the elemental Pd and Pt are relatively poor thermally conductive materials.

\begin{table*}[t]
\renewcommand{\arraystretch}{1.0} \setlength{\tabcolsep}{6pt}
\begin{center}
{%\footnotesize
\begin{tabular}{lccccccc} \hline \hline

\multicolumn{6}{c}{\bf{Alloying of metallic section of NFT and heat-sink for optimized nanoheating figures of merit}} \\[0.1cm] \hline

& \splitbox{$\Delta T_\mathrm{FePt}$\\ $/\Delta T_\mathrm{metal}$} & $T_\mathrm{metal}^\mathrm{max}$ (K) & $T_\mathrm{FePt}^\mathrm{max}$ (K) & \splitbox{Max. cross.\\track grad.\\(K / nm)} &  \splitbox{Max. down\\track grad.\\(K / nm)} \\ \cline{2-6}

(Au), $n\approx0.068$, $P_\mathrm{in}=3.75$~mW & \bf{6.2} & 389 & 891 & 16.7 & 17.6  \\  \cline{2-6} 

(AuAg\td{7})  $P_\mathrm{in}=4.00$~mW & \bf{6.5} & 381 & 867 & 15.3 & 16.0 \\  \cline{2-6}

(AuAg\td{3})  $P_\mathrm{in}=6.75$~mW & \bf{7.1} & 376 & 885 & 11.7 & 10.8  \\  \cline{2-6}

(Au\td{0.125}Ag\td{0.875})  $P_\mathrm{in}=4.50$~mW & \bf{7.0} & 381 & 911& 15.3 & 15.7 \\  \cline{2-6}

(Au\td{0.9}Pd\td{0.1}) $P_\mathrm{in}=6.50$~mW & \bf{6.2} & 393 & 911 & 11.3 & 15.4 \\  \cline{2-6}

\hline \\

\multicolumn{6}{c}{\bf{Alloying $10$~nm tip of metallic section}} \\[0.2cm] \hline

(Pd)  $P_\mathrm{in}=3.75$~mW & \bf{3.0} & 418 & 668 & 8.1 & 9.4 \\  \cline{2-6}

(Au\td{0.5}Pd\td{0.5}) $P_\mathrm{in}=3.75$~mW & \bf{4.5} & \splitbox{393 (436 in alloy section)} & 740 & 9.8 & 11.3 \\  \cline{2-6}

(Au\td{0.3}Pt\td{0.7}) $P_\mathrm{in}=4.25$~mW &\bf{4.7} & \splitbox{402 (432 in alloy section)} & 800 &  10.8 & 12.5 \\  \cline{2-6}

(Au\td{0.7}Pt\td{0.3}) $P_\mathrm{in}=4.25$~mW & \bf{4.8} & \splitbox{393 (402 in alloy section)}  & 775 & 9.4 & 11.1 \\

\hline \hline
\end{tabular}}
\end{center}
\caption{Simulated thermal efficiency of NFT with tapered, hybrid plasmonic waveguide~\cite{Abadia:18} using various Au-alloys. A temperature under 400 K is here considered  desirable for Au while maximizing the temperature in the media. High thermal gradients are also here considered  desirable in order to localize the heating as much as possible. Reported values for the refractive index of Au vary, and the value used here for Au of $n\approx0.068$ is derived from first principles.
}
\label{tab:therm-perf}
\end{table*}

We performed FEM simulations using the candidate alloys with promising plasmon generation and improved thermal conductivity.
In the NFT design, SPPs are generated by using a 3-fold process, namely; (1) effective index matching, for a photonic waveguide mode with the plasmonic mode, done by adjusting the thickness of the insulator layer, (2) designing the MIS waveguide component as a Fabry-Perot cavity, and (3) tapering to achieve localized heating suitable in media for HAMR. 
In Fig.~\ref{fig:hybrid-plas-wg}(b)~and~\ref{fig:hybrid-plas-wg}(c), we show the temperature, heat flux, and thermal gradients of the NFT and within the center of a FePt magnetic layer often used in recording media (see Supplemental Material~\cite{SeeSM} for full media materials/dimensions).
The size of the heated region may be controlled by adjusting the input power. 
This is vital for maximizing areal bit densities, i.e., cross (x-axis) and down (y) track regions, in magnetic recording devices. 
The bulk of the power within  the NFT is seen to be concentrated within the tip of the NFT (red region)~\cite{Bello:19}

Table~\ref{tab:therm-perf} contains a comparative analysis of the thermal performance~\cite{Datta2017} when replacing the Au components with various Au alloys as well as the last $10$~nm section from the ABS. 
Figure of merits used to quantify the thermal performance of the NFT are adopted from Refs.~\citenum{Datta2017}~and~\citenum{Abadia:18}. The dimensions of the NFT design used in simulations, being originally optimized using pure Au, are left unchanged.
Therefore, we notice comparable optical performance (required input power, P\td{in}) using alloys with similar refractive indices to Au. 
We find that in order to simultaneously have an improved thermal performance, alloys that have larger thermal conductivities than pure Au are desired. 
Therefore, Table~\ref{tab:therm-perf} reports on a number of Au\td{x}Ag\td{1−x} alloys used to replace the entire metallic section of the NFT with a noticeable effect. 
In particular, a crucial figure of merit for thermal stability, which compares the change in FePt temperature to that of the Au in the NFT ($\Delta$T\td{Media}/$\Delta$T\td{Au}), is markedly improved. 
It should be noted that thermal gradients in the recording layer, a crucial figure of merit for bit density, tends to be larger for Au alloys that index match well with pure Au and  more effectively localize power within the recording layer, i.e., more effectively demonstrate better optical efficiency. 

Au alloyed with a small percentage (10\%) of Pt/Pd, which are expected to have improved hardness by the rule of mixing, also performed well by demonstrating in simulation a similar figure of merit for thermal stability  to 
that of pure Au. 
However for alloys with higher concentrations of Pd/Pt, replacing the entire metallic component yielded a lower optical and thermally efficiency in simulation, due to poorer index matching between photonic and plasmonic modes. 
However, they were simulated to be potentially suitable for transducer operation when replacing a smaller section near the tip of the NFT where the majority of power is concentrated. 
Although optical and thermal efficiency was lower, these alloys are anticipated to have smaller coefficients of expansion. 

To conclude, we have calculated electronic and thermal material parameters used to measure the plasmonic performance for Au-M alloys suitable for nanoheating applications. By substituting, fully or partially, the metallic parts of a previously-introduced model NFT design, we demonstrate that plasmonic performance is generally maintained or reduced at $830$~nm. However, it potentially enables the engineering of  desired thermal performance of Au-based NFTs and possibly improved mechanical durability under relatively high-power conditions. 
We presented the multiscale design protocol as a proof-of-concept within a representative, previously-demonstrated NFT design; however, it is suitable to be modified for a number of nanoscale heating applications. 
Despite our focus on a single operation wavelength and temperature, our assessment procedure is easily 
applicable to any wavelength and reasonable temperature at which phonon contributions are relatively small.  
\section*{ACKNOWLEDGMENT}
We acknowledge the support of 
Trinity College Dublin School of Physics, 
Science Foundation Ireland (SFI)
through The Advanced Materials and Bioengineering Research Centre 
(AMBER, grant 12/RC/2278,  12/RC/2278\_P2 and 15/IA/2854), and  the European Regional Development Fund (ERDF) and the EU Horizon 2020 - Marie Sk\l odowska-Curie grant 713567.
We also acknowledge the DJEI/DES/SFI/HEA Irish Centre for High-End Computing (ICHEC) 
for the provision of computational facilities and support.
We further acknowledge Trinity Centre for High Performance Computing
and Science Foundation Ireland, for the maintenance and
funding, respectively, of the Boyle cluster on which
further calculations were performed.

%\bibliography{biblo}


\clearpage

%\begin{center}
%\textsc{Supplemental Material for}\\
%\textsc{``Multiscale Design of Au-Based Alloys for Improved Plasmon Delivery and Nanoheating in Near-Field Transducers''}
%\end{center}

\section*{Supplemental Material for\\
``Multiscale Design of Au-Based Alloys for Improved Plasmon Delivery and Nanoheating in Near-Field Transducers''}

\renewcommand\thefigure{SI\arabic{figure}} 
\setcounter{figure}{0} 
\setcounter{section}{0} 
\setcounter{table}{0} 
\setcounter{equation}{0} 
 
% \tableofcontents

\section{Computational methodology}
The initial crystallographic information for  bulk Au, Ag, Cu were adopted from Ref.~\citenum{Suh1988}, and for bulk Pd, and Pt from Ref.~\citenum{doi:10.1107/S0365110X65000361}. 
Bespoke, norm-conserving PBE pseudo-potentials for the elemental metals were produced using the pseudo-potential generator OPIUM~\cite{opium}, and the pre-processing tool within Quantum ESPRESSO software~\cite{0953-8984-21-39-395502,0953-8984-29-46-465901}  was used to produce the VCA pseudo-potentials. 
The approximate KS  electronic structures were obtained by performing full-geometry optimization, self-consistent field (SCF) and non-self-consistent field (NSCF) simulations, in order, using the  Quantum ESPRESSO software with the converged parameters  used from Ref.~\citenum{Orhan_2019}, as well as the  choice of smearing parameter for the Marzari-Vanderbilt cold smearing~\cite{PhysRevLett.82.3296}, namely $0.1$~eV. The \gw and spectral simulations for the inter-band contributions were performed using  the Yambo code~\cite{MARINI20091392}. 

The Fermi velocities, and Drude plasmon energies were evaluated using an in-house post-processing code starting from the KS band-structures~\cite{Orhan_2019}. 
For finite-element simulations, we have used scattering boundary conditions within the  COMSOL Multiphysics modeling  software for the outer boundaries of the cladding material (SiO\td{2})~\cite{Abadia:18,comsol}. 
The medium had a simulated in-plane area of $450  \times 750$~nm$^2$. 
Any resistive heating beyond this region was determined to have negligible impact on the area of maximum heating.
FEM simulations used a minimum element size of $4.2$~nm, maximum growth rate of $1.4$, and curvature factor of $0.4$ for convergence of results. We have only reported on the most optimal Au-based alloys identified in simulation 
for nanoscale heating with a specific antenna-based, plasmonic NFT.   

\section{Semi-empirical thermal parameters}
The thermal conductivity of metals is quite a complicated quantity, to which various scattering processes such as the electron-phonon, the electron-electron, as well as the electron-impurity scattering all contribute in the case of the noble metals and their alloys  up to their melting points~\cite{Uher2004}. 
The electrical thermal conductivity, $\kappa_\mathrm{e}$, is related to the electrical conductivity, $\sigma$,  via the Wiedermann-Franz law~\cite{doi:10.1002/andp.18531650802} given by

\begin{align}\label{eq:eq1}
\kappa_\mathrm{e}=\frac{3}{2}\left (\frac{k_\mathrm{B}}{e}\right)^2T \sigma,
\end{align}
where $k_\mathrm{b}$ and $T$ are the Boltzmann constant and  temperature, respectively. 
The electrical conductivity can be described within the Drude theory, which was used in Ref.~\citenum{Orhan_2019} to approximate the phenomenological plasmon life-times, $\tau_\mathrm{p}$, per
\begin{align}\label{eq:eq2}
\sigma=\frac{n_\mathrm{c} e^2 \tau_\mathrm{p}}{m_\mathrm{eff}},
\end{align}
where $n_\mathrm{c}$ is the carrier density, which is the  density of state at the Fermi level,  $N(E_\mathrm{F})$, per unit volume. 
The effective mass, $m_\mathrm{eff}$ can be approximated using the Fermi velocity, within the one-band theory assuming that the electronic bands have parabolic dispersion, by $m_\mathrm{eff} ^{-1}\approx \langle v^2(E_\mathrm{F})\rangle$.
Then, the electronic thermal conductivity is given by
\begin{align}\label{eq:eq3}
\kappa_\mathrm{e}=\frac{3}{2} k_\mathrm{B}^2 T \frac{N(E_\mathrm{F})}{V_\mathrm{cell}} \langle v^2(E_\mathrm{F})\rangle\tau_\mathrm{p}. 
\end{align}
For Au\td{x}M\td{1-x} alloys Eq.~\ref{eq:eq3} is reduces  to 
\begin{align}
\kappa_\mathrm{e}(x)=\left[ x f_\mathrm{Au}+ (1-x)f_\mathrm{M}\right] T  \frac{N(E_\mathrm{F})}{V_\mathrm{cell}}\langle v^2(E_\mathrm{F})\rangle ,
\end{align}
where the scaling factor $f_\mathrm{M}$ for the given elemental metal M is obtained from
\begin{align}
f_\mathrm{M}=\frac{\kappa_\mathrm{exp}^\mathrm{M}(T)}{T}\frac{V_\mathrm{cell}}{N(E_\mathrm{F})\langle v^2(E_\mathrm{F})\rangle },
\end{align} 
using the experimental thermal conductivity values at $400$~K for elemental metals shown below. 
\begin{table}[H]
\renewcommand{\arraystretch}{2} \setlength{\tabcolsep}{2pt}
\begin{center}
{%\scriptsize
\begin{tabular}{l|ccccc} \hline \hline 
 
& Au~\cite{Touloukian1970} & Ag~\cite{Touloukian1970} & Cu~\cite{Touloukian1970} & Pd~\cite{ref635953187} & Pt~\cite{ref9919} \\ \hline

\splitbox{$\kappa_\mathrm{exp}^\mathrm{M}$\\$T=400$~K}  & $312$ &$420$ & $392$ & $75$ & $72$ \\

\hline \hline
\end{tabular}}
\end{center}
\caption{Experimental thermal conductivity for the elemental metals at $T=400$~K used to determine the scaling parameters for alloys.}
\label{tab:au-cu}
\end{table}

The specific heat due to  charge carriers  is calculated using first-principles electronic parameters derived via the Drude definition of the electronic  heat capacity  per unit volume, $c_\mathrm{e}$, at low temperature, given by~\cite{Uher2004}
\begin{align}
C_\mathrm{e}=\frac{\pi^2}{3}N(E_\mathrm{F})E_\mathrm{F}k_\mathrm{B}\frac{T}{T_\mathrm{F}},
\end{align}
where $T_\mathrm{F}=E_\mathrm{F}/k_\mathrm{B}$ and $E_\mathrm{F}$ is the Fermi energy. 
Then the electronic specific heat can be expressed as 
\begin{align}
c_e=\frac{\pi^2}{3 m_\mathrm{avr}}N(E_\mathrm{F})k^2_\mathrm{B}T,
\end{align}
where $m_\mathrm{avr}$ is the averaged atomic mass for  the Au\td{x}M\td{1-x}  given by
\begin{align}
m_\mathrm{avr}=x m_\mathrm{Au}+(1-x)m_\mathrm{M}.
\end{align}

\section{Optimized first-principles simulation parameters}
\begin{table}[H]
\renewcommand{\arraystretch}{1.3} \setlength{\tabcolsep}{10pt}
\begin{center}
{%\footnotesize
\begin{tabular}{llc} \hline \hline 
Level  & Parameter &  Value   \\ \hline
\multirow{2}{*}{GeoOpt} & \splitbox{The plane-wave kinetic\\energy cutoff energies}  & $75$~Ry \\
&k-points  & $4\times 4 \times 4$ \\ \hline
\multirow{2}{*}{SCF}& \splitbox{The plane-wave kinetic\\energy cutoff energies}  & $25$~Ry \\
&k-points  &  $21\times 21 \times 21$\\ \hline
\multirow{3}{*}{NSCF}&\splitbox{The plane-wave kinetic\\energy cutoff energies} & $25$~Ry \\
&k-points  &  $16\times 16 \times 16$\\ 
& Number of bands & 100 \\ \hline
\multirow{2}{*}{\gw}& \splitbox{Number of bands\\in the self-energy exchange}  &10 \\
& \splitbox{Number of bands\\in the self-energy correlation}  &100 \\
\hline \hline
\end{tabular}}
\end{center}
\caption{The converged parameters used in the geometry optimization (GeoOpt),  self-consistent field (SCF), non-self-consistent field (NSCF), and \gw.}
\label{tab:dft-par}
\end{table}

\section{Thermal and optical parameters}

\begin{table}[H]
\renewcommand{\arraystretch}{1.3} \setlength{\tabcolsep}{6pt}
\begin{center}
{\footnotesize
\begin{tabular}{lcccc} \hline \hline 
System & \splitbox{ $\kappa$ \\(W m$^{-1}$ K$^{-1}$)} &  \splitbox{ $c_\mathrm{e}$ \\($\times 10^{-3}$\\J gm$^{-1}$ K$^{-1}$)}  & n & k \\ \hline
 
Au                &  312.0  &  0.273  &  0.068  &  5.176 \\
Au$_7$Ag          &  271.2  &  0.297  &  0.148  &  5.454\\
Au$_3$Ag          &  600.0  &  0.326  &  0.115  &  8.133\\
AuAg              &  467.2  &  0.329  &  0.071  &  6.467\\
AuAg$_3$          &  669.2  &  0.417  &  0.091  &  8.034\\
AuAg$_7$          &  347.9  &  0.452  &  0.054  &  5.485\\
Ag                &  420.0  &  0.459  &  0.045  &  5.675\\

\hline \hline
\end{tabular}}
\end{center}
\caption{The thermal conductivity ($\kappa$), and the electronic specific heat ($c_\mathrm{e}$) at $400$~K; the refractive index ($n$), and the extinction coefficient ($k$) at $830$~nm  of the mono-layer-stacked ordered Au-Ag alloys for available stoichiometric ratios.}
\label{tab:au-ag-ordered}
\end{table}

\begin{table}[H]
\renewcommand{\arraystretch}{1.3} \setlength{\tabcolsep}{4pt}
\begin{center}
{\footnotesize
\begin{tabular}{lcccc} \hline \hline 
System & \splitbox{ $\kappa$ \\(W m$^{-1}$ K$^{-1}$)} &  \splitbox{ $c_\mathrm{e}$ \\($\times 10^{-3}$\\J gm$^{-1}$ K$^{-1}$)}  & n & k \\ \hline

Au                       &  312.0  &  0.273  &  0.068  &  5.176 \\
Au\td{0.875}Ag\td{0.125} &  340.9  &  0.265  &  0.077  &  5.207\\
Au\td{0.75}Ag\td{0.25}   &  365.0  &  0.281  &  0.072  &  5.342\\
Au\td{0.5}Ag\td{0.5}     &  390.1  &  0.306  &  0.057  &  5.545\\
Au\td{0.25}Ag\td{0.75}   &  465.7  &  0.388  &  0.057  &  6.067\\
Au\td{0.125}Ag\td{0.875} &  469.0  &  0.413  &  0.053  &  6.096\\
Ag                       &  420.0  &  0.459  &  0.045  &  5.675\\

\hline \hline
\end{tabular}}
\end{center}
\caption{The thermal conductivity ($\kappa$), and the electronic specific heat ($c_\mathrm{e}$) at $400$~K; the refractive index ($n$), and the extinction coefficient ($k$) at $830$~nm  of the disordered Au\td{x}Ag\td{1-x} alloys for available stoichiometric ratios.}
\label{tab:au-ag-vca}
\end{table}

\begin{table}[H]
\renewcommand{\arraystretch}{1.3} \setlength{\tabcolsep}{6pt}
\begin{center}
{\footnotesize
\begin{tabular}{lcccc} \hline \hline 
System & \splitbox{ $\kappa$ \\(W m$^{-1}$ K$^{-1}$)} &  \splitbox{ $c_\mathrm{e}$ \\($\times 10^{-3}$\\J gm$^{-1}$ K$^{-1}$)}  & n & k \\ \hline
 
Au                  &  312.0  &  0.273  &  0.068  &  5.176 \\
Au\td{7}Cu          &  253.9  &  0.329  &  0.159  &  5.295\\
Au\td{3}Cu          &  591.1  &  0.373  &  0.140 &  8.194\\
AuCu                &  451.3  &  0.570  &  0.110  &  6.629\\
AuCu\td{3}          &  586.7  &  0.775  &  0.134  &  7.732\\
AuCu\td{7}          &  278.9  &  0.939  &  0.123  &  4.969\\
Cu                  &  392.0  &  1.199  &  0.080  &  5.397\\

\hline \hline
\end{tabular}}
\end{center}
\caption{The thermal conductivity ($\kappa$), and the electronic specific heat ($c_\mathrm{e}$) at $400$~K; the refractive index ($n$), and the extinction coefficient ($k$) at $830$~nm  of the mono-layer-stacked ordered Au-Cu alloys for available stoichiometric ratios.}
\label{tab:au-cu-ordered}
\end{table}

\begin{table}[H]
\renewcommand{\arraystretch}{1.3} \setlength{\tabcolsep}{4pt}
\begin{center}
{\footnotesize
\begin{tabular}{lcccc} \hline \hline 
System & \splitbox{ $\kappa$ \\(W m$^{-1}$ K$^{-1}$)} &  \splitbox{ $c_\mathrm{e}$ \\($\times 10^{-3}$\\J gm$^{-1}$ K$^{-1}$)}  & n & k \\ \hline

Au                       &  312.0  &  0.273  &  0.068  &  5.176\\
Au\td{0.875}Cu\td{0.125} &  341.4  &  0.313  &  0.094  &  5.227\\
Au\td{0.75}Cu\td{0.25}   &  365.6  &  0.386  &  0.109  &  5.186\\
Au\td{0.5}Cu\td{0.5}     &  387.8  &  0.572  &  0.128  &  5.480\\
Au\td{0.25}Cu\td{0.75}   &  395.0  &  0.858  &  0.129  &  5.327\\
Au\td{0.125}Cu\td{0.875} &  379.2  &  1.012  &  0.130  &  4.916\\
Cu                       &  392.0  &  1.199  &  0.080  &  5.397\\

\hline \hline
\end{tabular}}
\end{center}
\caption{The thermal conductivity ($\kappa$), and the electronic specific heat ($c_\mathrm{e}$) at $400$~K; the refractive index ($n$), and the extinction coefficient ($k$) at $830$~nm  of the disordered Au\td{x}Cu\td{1-x} alloys for available stoichiometric ratios.}
\label{tab:au-cu-vca}
\end{table}

\begin{table}[H]
\renewcommand{\arraystretch}{1.3} \setlength{\tabcolsep}{4pt}
\begin{center}
{\footnotesize
\begin{tabular}{lcccc} \hline \hline 
System & \splitbox{ $\kappa$ \\(W m$^{-1}$ K$^{-1}$)} &  \splitbox{ $c_\mathrm{e}$ \\($\times 10^{-3}$\\J gm$^{-1}$ K$^{-1}$)}  & n & k \\ \hline

Au                   &  312.0  &  0.273  &  0.068  &  5.176\\
Au\td{0.9}Pd\td{0.1} &  488.5  &  0.330  &  0.098  &  7.560\\
Au\td{0.8}Pd\td{0.2} &  265.9  &  0.219  &  0.173  &  6.632\\
Au\td{0.7}Pd\td{0.3} &  619.4  &  0.611  &  0.500  &  8.861\\
Au\td{0.6}Pd\td{0.4} &  468.8  &  0.479  &  0.518  &  7.675\\
Au\td{0.5}Pd\td{0.5} &  143.7  &  0.949  &  1.274  &  5.505\\
Au\td{0.4}Pd\td{0.6} &  211.1  &  1.997  &  1.275  &  6.113\\
Au\td{0.3}Pd\td{0.7} &  220.7  &  2.437  &  1.131  &  6.942\\
Au\td{0.2}Pd\td{0.8} &  204.8  &  2.838  &  1.137  &  7.120\\
Au\td{0.1}Pd\td{0.9} &  179.3  &  3.364  &  1.219  &  7.115\\
Pd                   &  75.0  &  4.812  &  1.544  &  3.954\\

\hline \hline
\end{tabular}}
\end{center}
\caption{The thermal conductivity ($\kappa$), and the electronic specific heat ($c_\mathrm{e}$) at $400$~K; the refractive index ($n$), and the extinction coefficient ($k$) at $830$~nm  of the disordered Au\td{x}Pd\td{1-x} alloys for available stoichiometric ratios.}
\label{tab:au-pd-vca}
\end{table}

\begin{table}[H]
\renewcommand{\arraystretch}{1.3} \setlength{\tabcolsep}{4pt}
\begin{center}
{\footnotesize
\begin{tabular}{lcccc} \hline \hline 
System & \splitbox{ $\kappa$ \\(W m$^{-1}$ K$^{-1}$)} &  \splitbox{ $c_\mathrm{e}$ \\($\times 10^{-3}$\\J gm$^{-1}$ K$^{-1}$)}  & n & k \\ \hline

Au                   &  312.0  &  0.273  &  0.068  &  5.176\\
Au\td{0.9}Pt\td{0.1} &  493.5  &  0.329  &  0.101  &  7.385\\
Au\td{0.8}Pt\td{0.2} &  303.2  &  0.261  &  0.670  &  6.879\\
Au\td{0.7}Pt\td{0.3} &  602.0  &  0.586  &  0.908  &  9.027\\
Au\td{0.6}Pt\td{0.4} &  477.6  &  0.529  &  0.957  &  8.721\\
Au\td{0.5}Pt\td{0.5} &  143.3  &  0.670  &  1.177  &  5.571\\
Au\td{0.4}Pt\td{0.6} &  183.5  &  1.030  &  1.091  &  6.565\\
Au\td{0.3}Pt\td{0.7} &  179.1  &  1.374  &  1.211  &  6.448\\
Au\td{0.2}Pt\td{0.8} &  184.1  &  1.696  &  1.033  &  7.029\\
Au\td{0.1}Pt\td{0.9} &  127.4  &  2.110  &  1.815  &  5.382\\
Pt                   &  72.0  &  2.150  &  1.363  &  5.406\\

\hline \hline
\end{tabular}}
\end{center}
\caption{The thermal conductivity ($\kappa$), and the electronic specific heat ($c_\mathrm{e}$) at $400$~K; the refractive index ($n$), and the extinction coefficient ($k$) at $830$~nm  of the disordered Au\td{x}Pt\td{1-x} alloys for available stoichiometric ratios.}
\label{tab:au-pt-vca}
\end{table}

\section{Finite-element simulation parameters}

\begin{table}[h]
\renewcommand{\arraystretch}{1.3} \setlength{\tabcolsep}{15pt}
\begin{center}
{\footnotesize
\begin{tabular}{lc} \hline \hline 
\multicolumn{2}{c}{Waveguide dimensions} \\ \hline

Si waveguide length, width ($x$) & 0.5~$\mu$m, 450~nm \\
Si waveguide thickness ($y$) & 250~nm\\
Au film thickness & 60~nm \\
Taper /Au/SiO\td{2}/Si length ($z$) & 330~nm \\
Au, SiO\td{2}, Si tip width & 20~nm \\
SiO\td{2} thickness & 10~nm \\
Heat-sink (Au) length, width & 222~nm, 450~nm \\
Heat-sink thickness & 150~nm \\
Write pole length, width & 50~nm, 450~nm \\
Write pole thickness & 100~nm \\
\hline \hline
\end{tabular}}
\end{center}
\caption{Dimensions from Refs.~\cite{Bello:20,Abadia:18} of the photonic and plasmonic waveguide design used as
an example in simulations.}
\label{tab:au-pt-vca}
\end{table}

\newpage

\onecolumngrid

\begin{table}[H]
\renewcommand{\arraystretch}{1.3} \setlength{\tabcolsep}{5pt}
\begin{center}
{\footnotesize
\begin{tabular}{lcccc} \hline \hline 

\multicolumn{5}{c}{Film}\\ \hline
& Thickness (nm) & \splitbox{$n$\\$\lambda=830$~nm} & \splitbox{$C$\\( $\times 10^6$J m\tu{-3}K\tu{-1})} & \splitbox{$\kappa$\\(W m\td{-1} K\tu{-1})} \\ \hline

Heated air/lubricant &  2.5 & 1.0 & 1.0& 3.0 \\

Carbon overcoat & 6.5 &  2.4+0.51\it{i} & 1.0 & 1.0 \\

Capping & 1.5 nm & 4.71+3.38\it{i} & 2.89 & 20.0 \\

FePt & 9 & 3.04+2.69\it{i} & 2.89 & 5.7, 1.4 (in-plane) \\

MgO & 7 & 1.65 & 3.36 & 30.0 \\
Amorphous underlayer & 9 & 5.87+4.68\it{i} & 1.1 & 50.0 \\
Heat-sink & 72 & 3.72+3.99\it{i} & 3.06 & 10.5 \\ \hline \\[-0.2cm]

\multicolumn{5}{c}{Materials (NFT, waveguide)}\\ \hline
Au &   & 0.16+5.15\it{i} & 2.49 & 317 \\
SiO\td{2} & 1.45 & 1.6 & 1.4 \\
Si & & 3.67 & 1.63 & 131 \\
Write pole & &2.14+4.24\it{i} & 3.73 & 3.0 \\  
\hline \hline
\end{tabular}}
\end{center}
\caption{Film thickness,  refractive index ($n$) at $830$~nm,   specific heat capacity ($C$), and   electronic thermal conductivity ($\kappa$) values from Refs.~\cite{Bello:20,Abadia:18}, 
used here as an example for  finite-element simulations.}

\label{tab:au-pt-vca}
\end{table}

\twocolumngrid

%\newpage
%\bibliography{../Main/biblo}


\end{document}